\documentclass[prb,preprint,aps,superscriptaddress]{revtex4-1}
\usepackage{epsfig}
\usepackage{amssymb}
\usepackage{amsmath}
\usepackage{mathrsfs}
\usepackage{amsfonts}
\usepackage{threeparttable}
\usepackage{graphicx}
\usepackage{natbib}
\usepackage{longtable}
\usepackage{txfonts}
\usepackage{color}
\usepackage{epstopdf}
\usepackage{braket}
\usepackage{lineno}
\usepackage{xr}
\begin{document}
\title{Non-Hermitian Optical Parametric Systems with Anti-parity-time Symmetry}
\author{Ben Li}
\thanks{These authors contributed equally to this work.}
\affiliation{School of Information Science and Technology and Key Laboratory for Information Science of Electromagnetic Waves (MOE) and Department of Optical Science and Engineering and Key Laboratory of Micro and Nano Photonic Structures (MOE), Fudan University, Shanghai 200433, China}

\author{Yanfang Zhang}	
\thanks{These authors contributed equally to this work.}
\affiliation{Key Laboratory for Laser Plasma, School of Physics and Astronomy, Shanghai Jiao Tong University, Shanghai 200240, China}

\author{Jing Wang}	
\email{wangj1118@sjtu.edu.cn}
\affiliation{Key Laboratory for Laser Plasma, School of Physics and Astronomy, Shanghai Jiao Tong University, Shanghai 200240, China}

\author{Wenhao Wang}	
\affiliation{Key Laboratory for Laser Plasma, School of Physics and Astronomy, Shanghai Jiao Tong University, Shanghai 200240, China}

\author{Jingui Ma}	
\affiliation{Key Laboratory for Laser Plasma, School of Physics and Astronomy, Shanghai Jiao Tong University, Shanghai 200240, China}

\author{Peng Yuan}	
\affiliation{Key Laboratory for Laser Plasma, School of Physics and Astronomy, Shanghai Jiao Tong University, Shanghai 200240, China}

\author{Dongfang Zhang}	
\affiliation{Key Laboratory for Laser Plasma, School of Physics and Astronomy, Shanghai Jiao Tong University, Shanghai 200240, China}

\author{Yongfeng Mei}
\affiliation{Department of Materials Science, Fudan University, Shanghai 200433, China}

\author {Heyuan Zhu}
\affiliation{School of Information Science and Technology and Key Laboratory for Information Science of Electromagnetic Waves (MOE) and Department of Optical Science and Engineering and Key Laboratory of Micro and Nano Photonic Structures (MOE), Fudan University, Shanghai 200433, China}

\author{Hao Zhang} \email{zhangh@fudan.edu.cn}
\affiliation{School of Information Science and Technology and Key Laboratory for Information Science of Electromagnetic Waves (MOE) and Department of Optical Science and Engineering and Key Laboratory of Micro and Nano Photonic Structures (MOE), Fudan University, Shanghai 200433, China}
\affiliation{Yiwu Research Institute of Fudan University, Chengbei Road, Yiwu City, Zhejiang 322000, China}

\author{Liejia Qian}	
\affiliation{Key Laboratory for Laser Plasma, School of Physics and Astronomy, Shanghai Jiao Tong University, Shanghai 200240, China}
\affiliation{Tsung-Dao Lee Institute, Shanghai Jiao Tong University, Shanghai 200240, China}

\begin{abstract}
The continuous advancements in ultrafast lasers, characterized by high pulse energy, great average power, and ultrashort pulse duration, have opened up new frontiers and applications in various fields such as high-energy-density science. In this study, we investigated the implementation of non-Hermitian nonlinear parametric amplification by introducing anti-parity-time (anti-PT) symmetry to three-wave interaction processes. By exploring the parameter space defined by the coupling coefficient, phase mismatch, and absorption, we categorized the behavior of the non-Hermitian optical parametric system into four distinct quadrants, representing unbroken/broken anti-PT symmetry and amplification/attenuation, and amplification-attenuation boundaries and exceptional lines can be observed in such parametric space. Through simulations of the dynamical behavior of the interacting waves, we demonstrated the rich evolutions of the signal and idler waves in systems belonging to the respective quadrants and near exceptional points, revealed by the unique performance of eigenmodes. Our findings provide insights into the evaluation of energy flow direction in optical parametric amplification engineering by the directly linked parameter space, which contribute to a deeper understanding of photonics and laser science, potentially leading to new applications in these fields.
\end{abstract}

\maketitle
 
\textit{Introduction}--A significant application of nonlinear optics is the nonlinear frequency conversion via parametric wave-mixing processes, which are indispensable tools for building new light sources. Although several new records regarding pulse duration, intensity, pulse energy, average power and so on have been achieved in the past decades\cite{Dubietis1992, Cerullo2003, Vaupel2013}, the intrinsic limiting factors in optical parametric amplification (OPA) processes set the major obstacles to further increase the amplifier efficiency. In the OPA processes, the pump photon is splitted into one signal photon and the dual idler photon, and the cyclic exchange of pump and signal/idler photons known as the conversion-back-conversion cycles may happen when the spatiotemporal asynchronicity in the conversion processes exists. These limiting factors lead to the poor amplifier efficiency less than 20\% in early experiments\cite{Herrmann2009, Yunpei2012, Giedrius2011}. 

Recently, variants of OPAs that rely on dissipative loss of idler photons have been proposed and employed, which prevent energy return to pump by breaking the duality of signal and idler photons through eliminating idler photons, requiring the energy exchange between OPA systems and environments via idler photons. The pioneering work regarding the dissipative schemes was the quasi-parametric amplification (QPA) method, in which the idler is dissipatively depleted in the new borate crystal yttrium calcium oxyborate (YCOB) with samarium doping, and a pump depletion of 85$\%$ and a signal efficiency of 55$\%$ were achieved in the typical pump case using a spatiotemporal Gaussian laser\cite{Ma2015,Ma2017,Ma2022}. Later, the hybridization of parametric amplification with idler second-harmonic generation (SHG) to induce unusual evolution dynamics for a fully parametric amplifier, was proposed and has achieved a 48 dB gain with 68$\%$ quantum efficiency and 44$\%$ pump-to-signal energy conversion\cite{Flemens2021, Flemens2022,Flemens2022a}. It is worth noting that the dissipative schemes, irrespective of the introduction of material losses or additional nonlinear interactions, essentially create an open non-Hermitian system in the OPA processes. Actually, the non-Hermitian physical systems have received great attention due to their novel characteristics different from closed Hermitian systems\cite{ElGanainy2018, Oezdemir2019, ElGanainy2019}, transferring completely new regulation mechanisms for classical photon\cite{Weimann2017, Zhao2018}, phonon\cite{Fleury2015, Gao2021}, circuit\cite{Helbig2020} and mechanical\cite{Coulais2017, Ghatak2020} systems. Numerous facility properties containing single-mode lasing\cite{Zhao2018, Hodaei2014}, unidirectional invisibility caused by skin effect\cite{Lin2011, Lv2017}, asymmetric mode conversion\cite{Ghosh2016, Doppler2016}, exceptional point improved sensitivity\cite{Hodaei2017, Chen2017} and enhanced efficiency and bandwidth of parametric amplification\cite{ElGanainy2015, Antonosyan2015} have been proposed and realized depending on conscientious engineering schemes of incoherent gain and loss. 

In this work, inspired by the achievement of QPA method and the progress of non-Hermitian physics, we investigate a particular implementation of non-Hermitian parametric processes based on anti-PT symmetry in a single crystal where the signal and idler components may experience optical losses. From the perspective of non-Hermitian optics, it is the coherent coupling of dissipation, pumping and phase mismatch that determines the direction of energy flow, and is characterized by high efficiency and broadband potential in practical devices. We demonstrate the quadrant diagram of anti-PT symmetry and amplification parameters, and simulate the dynamics of amplified and reduced modes, finding that amplification can exist in both broken and unbroken anti-PT symmetry phases and the eigenmodes display rich behaviors.

In quadratic nonlinear crystals, under slowly varying envelope and undepleted pump approximations, the coupled-mode equations describing the parametric processes for signal and idler waves can be written in a simplified Hamiltonian form as shown in Eqs.~(\ref{equation:s1}-\ref{equation:s4}), which can be further decomposed by the Pauli matrices as

\begin{equation}
\mathcal{H}=\frac{1}{2} \left[ - i \gamma_{+}\tau_0 +  \kappa_{-} \tau_x - i \kappa_{+} \tau_y +  \left(\beta + i \gamma_{-}  \right) \tau_z \right]
\label{equation:1}
\end{equation}

where $\gamma_{\pm}$ is real with $\gamma_{\pm} = \gamma_i\pm\gamma_s$ and $\kappa_{\pm}$ is complex with $\kappa_{\pm} = \kappa_i E_p^* \pm \kappa_s E_p$, representing the losses and coupling of the system, $\tau_{x,y,z}$ are the Pauli matrices and $\tau_0$ is the $2\times2$ identity matrix. The Hamiltonian of Eq.~(\ref{equation:1}) is essentially the one for generalized two-level systems as $\mathcal{H}=a_0\tau_0+\mathbf{a(z)}\cdot\mathbf{\tau}$, with $a_0=-i\gamma_{+}/2$ and $\mathbf{a(z)}=[\kappa_{-}/2,-i\kappa_{+}/2,(\beta+i\gamma_{-})/2]$. The eigenvalues for Eq.~(\ref{equation:1}) are $eigs(\mathcal{H})=a_0\pm |\mathbf{a(z)}|$. Here, the real parts of $eigs(\mathcal{H})$ define the phase velocity, while the imaginary parts of $eigs(\mathcal{H})$ determine the modal gain or loss. When $\gamma_s,\gamma_i\ne0$ or $\kappa_i\ne-\kappa_s$, the Hamiltonian as shown in Eq.~(\ref{equation:1}) is no longer Hermitian, which indicates that, the non-trivial absorptions or couplings to the pump for the signal and idler photons will definitely drive the coherent three-mode Hermitian nonlinear processes into a non-Hermitian regime, in which the energy and probability conservation no longer persist under this situation. For bosons systems such as the optical parametric processes, the parity operator $P$ and time reversal operator $T$ can be defined as $P=\tau_x$ and $T=K$ (the complex conjugation operator), and the Hamiltonian can be invariant imposed under the PT symmetry, i.e. $[PT,\mathcal{H}]=0$, or anti-PT symmetry, i.e. $\{PT,\mathcal{H}\}\equiv PT\mathcal{H}+\mathcal{H}PT=0$, with $P^2=1$, $T^2=1$. The interaction Hamiltonian in Eq.~(\ref{equation:1}) obeys PT symmetry when $\gamma_s=-\gamma_i$, indicating that one can create a non-Hermitian optical parametric system with PT symmetry by imposing an anti-symmetric gain/loss profile to the signal/idler photons, which is analogous to the configurations widely studied in non-Hermitian photonics\cite{ElGanainy2019}. However, the interaction Hamiltonian in Eq.~(\ref{equation:1}) obeys the anti-PT symmetry when $\gamma_s=\gamma_i$, indicating that the signal and idler photons experience symmetric absorptions. Since the anti-PT operation leads to the self-collapse of eigenmodes of the Hamiltonian in Eq.~(\ref{equation:1}), and thus $eig(\mathcal{H})$ equals to its opposite conjugation, i.e. $eig(\mathcal{H})=-eig(\mathcal{H})^*$, indicating the eigenvalues of anti-PT-symmetric Hamiltonian are pure imaginary numbers, which is distinct from the systems with PT symmetry requiring $eigs(\mathcal{H})$ real numbers.

\begin{figure}[htbp]
\centering
\includegraphics[width=1\linewidth]{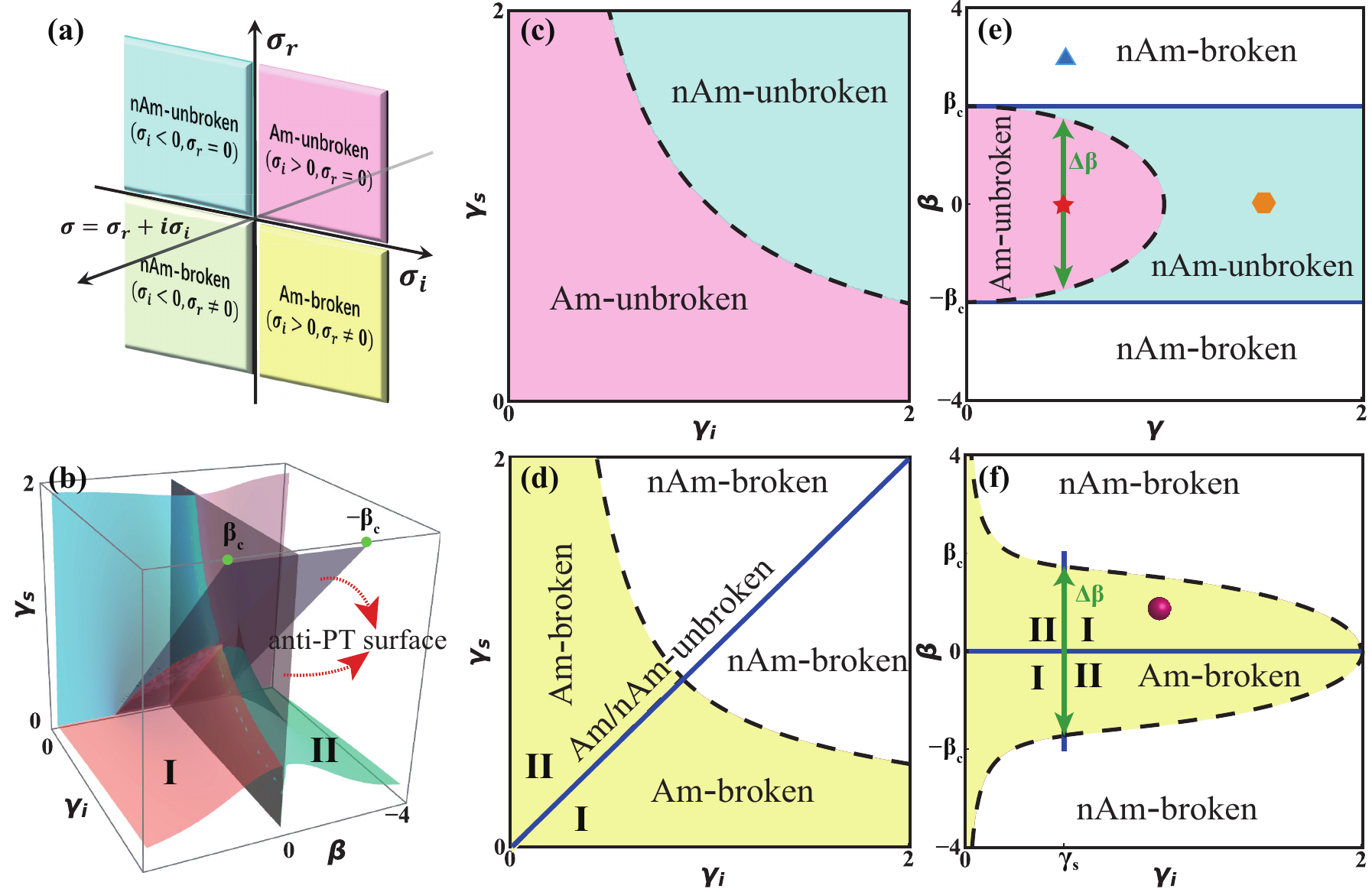}
\caption{(a) Quadrant diagrams and (b) parameter space of anti-PT symmetry and amplification. In (b), the blue and pink sheets represent the boundary of amplification for the two eigenmodes, and the gray sheet is the anti-PT symmetric surface. (c-f) The region of amplification/decay and unbroken/broken anti-PT symmetry considering (c) $\beta=0$, (d) $\beta=1$, (e) $\gamma_s=\gamma_i$ and (f) $\gamma_s=0.5$. The black dashed and blue solid lines represent the amplification-decay boundary and anti-PT-symmetry broken-unbroken boundary. For all the plots, $\kappa_s=\kappa_i=1$ and $|E_p|=1$. }
\label{fig:1}
\end{figure}

\textit{Full landscape of anti-PT symmetry and amplification in OPA}--For OPA with dissipative signal and idler channels, the restricted conditions for anti-PT symmetry are 1) perfect phase-matching $\beta=0$; 2) symmetric absorption for signal and idler channels ($\gamma_s=\gamma_i=\gamma$) and that the phase mismatch is less than a critical value $\beta_c=\sqrt{\kappa_+^2-\kappa_-^2}$. By intuition, the behaviors of the non-Hermitian parametric down-conversion processes can be categorized into four quadrants regarding the anti-PT symmetry and amplification, as shown in Fig.~\ref{fig:1}(a). For open systems with symmetric absorptions and small phase mismatches ($|\beta|<\beta_c$), which obey the anti-PT symmetry, amplification occurs if the absorptions $\gamma$ for signal and idler are smaller than the threshold value, $\frac{1}{2}\sqrt{\beta_c^2-\beta^2}$, in the first quadrant, but if the absorptions are larger than the threshold, decay occurs, as shown in the second quadrant. The open systems are anti-PT-symmetry-broken when $\gamma_s\ne\gamma_i$ or $|\beta|> \beta_c$, in which the signal beams can be still amplified if $\rm{Im}(\sigma)=\rm{Im}\left [ \sqrt{(\beta+i\gamma_-)^2-\beta_c^2} \right ] -\gamma_+>0$, shown in the fourth quadrant. The third quadrant corresponds to the anti-PT-symmetry-broken systems with negative imaginary parts of eigenvalues. The situations for anti-PT symmetry and amplification/decay in the parameter space including phase mismatch $\beta$ and absorptions for signal/idler channels $\gamma_s/\gamma_i$ is shown in Fig.~\ref{fig:1}(b), which manifests a large part of such parameter space correspond to amplification regime.

Specifically, for phase-matching systems obeying the anti-PT symmetry as shown in Fig.~\ref{fig:1}(c), the parameter boundary separating amplification and decay regimes called as amplification-decay boundary is $\gamma_s\gamma_i=\frac{1}{4}\beta_c^2$. Since the duality of signal and idler photons is preserved by anti-PT symmetry, symmetric amplification for the absorptions of signal/idler beams can be thus observed. For lossful  systems with a fixed phase mismatch smaller than $\beta_c$ as shown in Fig.~\ref{fig:1}(d), in the anti-PT-symmetry broken region other than the diagonal line, the amplification/decay is no longer symmetric regarding the absorptions of signal/idler channels for the respective eigenmode, but similarly, the amplification-decay boundary denoted as the dashed line can be observed therein, determined by the sign of $\rm{Im}(\sigma)$ as above-mentioned. For systems with symmetric absorptions for signal and idler channels ($\gamma_s=\gamma_i=\gamma$), the conditions for phase mismatch for anti-PT symmetry and amplification are $|\beta|<\beta_c$ and $|\beta|<\sqrt{\beta_c^2-4\gamma^2}$, implying amplification merely occurs in anti-PT symmetric region with the parameter boundary of $|\beta|=\sqrt{\beta_c^2-4\gamma^2}$, as shown in Fig.~\ref{fig:1}(e). However, when the symmetric absorption condition is relaxed by fixing $\gamma_s$ as shown in Fig.~\ref{fig:1}(f), the systems along the $\beta=0$ and $\gamma_i=\gamma_s$ lines are anti-PT symmetric, and the amplification-decay boundary can be determined by the sign of $\rm{Im}(\sigma)$ similarly. 

\begin{figure}[ht!]
	\centering
	\includegraphics[width=1\linewidth]{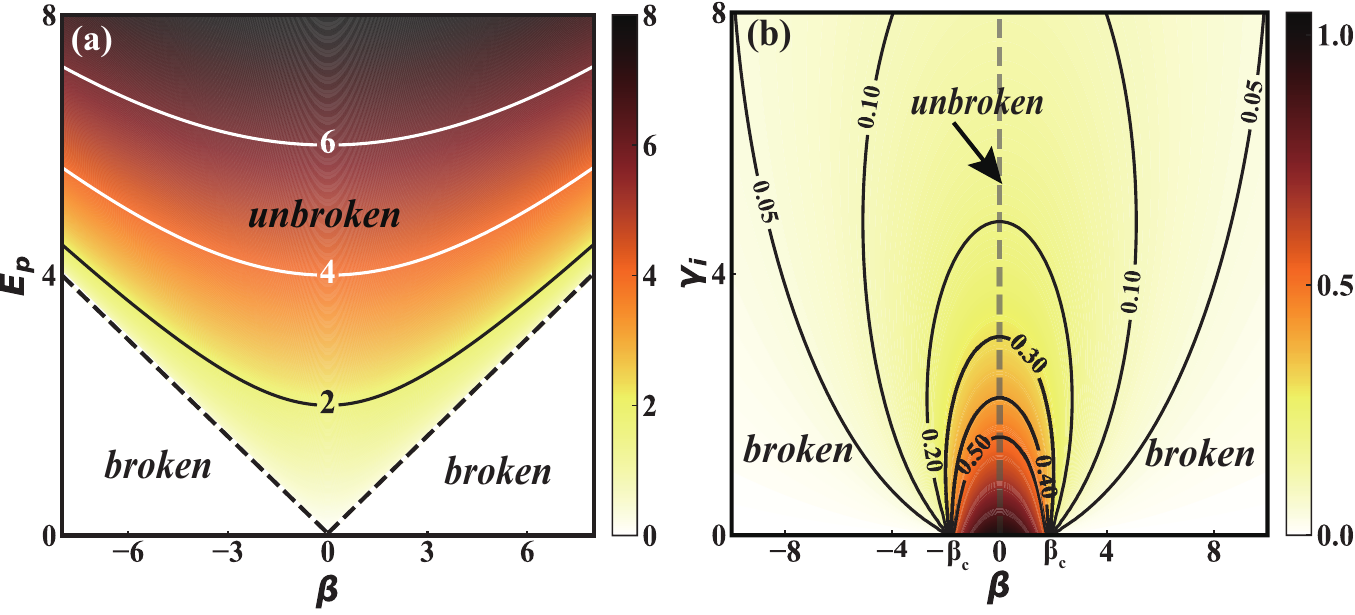}
	\caption{Largest mode amplification with (a) non-loss, $\gamma_i=0$, and (b) only considering idler's loss, $\gamma_i=0.2$, and $\gamma_s=0$, $\kappa_s=\kappa_i=1$ and $E_p=1$.}
	\label{fig:2}
\end{figure}

\textit{Impact of dissipative idler channel}--The phase-mismatch $\beta$-dependent largest mode amplification $Max[eigs(\mathcal{H})]$ for conventional OPA and the assumption for the QPA scheme is calculated and shown in Fig.~\ref{fig:2}, in which the solid lines stand for the relative values of gain (Im$(\sigma)$), while the colored and white regions denote the anti-PT symmetric unbroken and broken regimes, respectively. It is observed that the parametric amplification for dissipationless closed system can be realized only in anti-PT symmetric regime, and the amplification increases with the increasing pump intensity. For open systems with $\gamma_s=0$ and $\gamma_i\neq0$ (QPA), the anti-PT symmetry occurs only at perfect phase matching, i.e. $\beta=0$, and all the parametric down-conversion processes with non-trivial phase mismatch are anti-PT symmetry-broken with amplified signal beams. As shown in Fig.~\ref{fig:2}(b), when the phase mismatch $|\beta|\le\beta_c$, the gain decreases but the allowed $\beta$ region ($\Delta\beta$) rises with the increase of idler's loss, indicating the competition relation between gain and $\Delta\beta$. However, if $\beta>\beta_c$, the trend is opposite and the gain is much smaller than the corresponding $\beta<\beta_c$ case. Furthermore, smaller idler's loss and phase mismatch lead to larger gain.

\begin{figure}[t]
	\centering
	\includegraphics[width=1\linewidth]{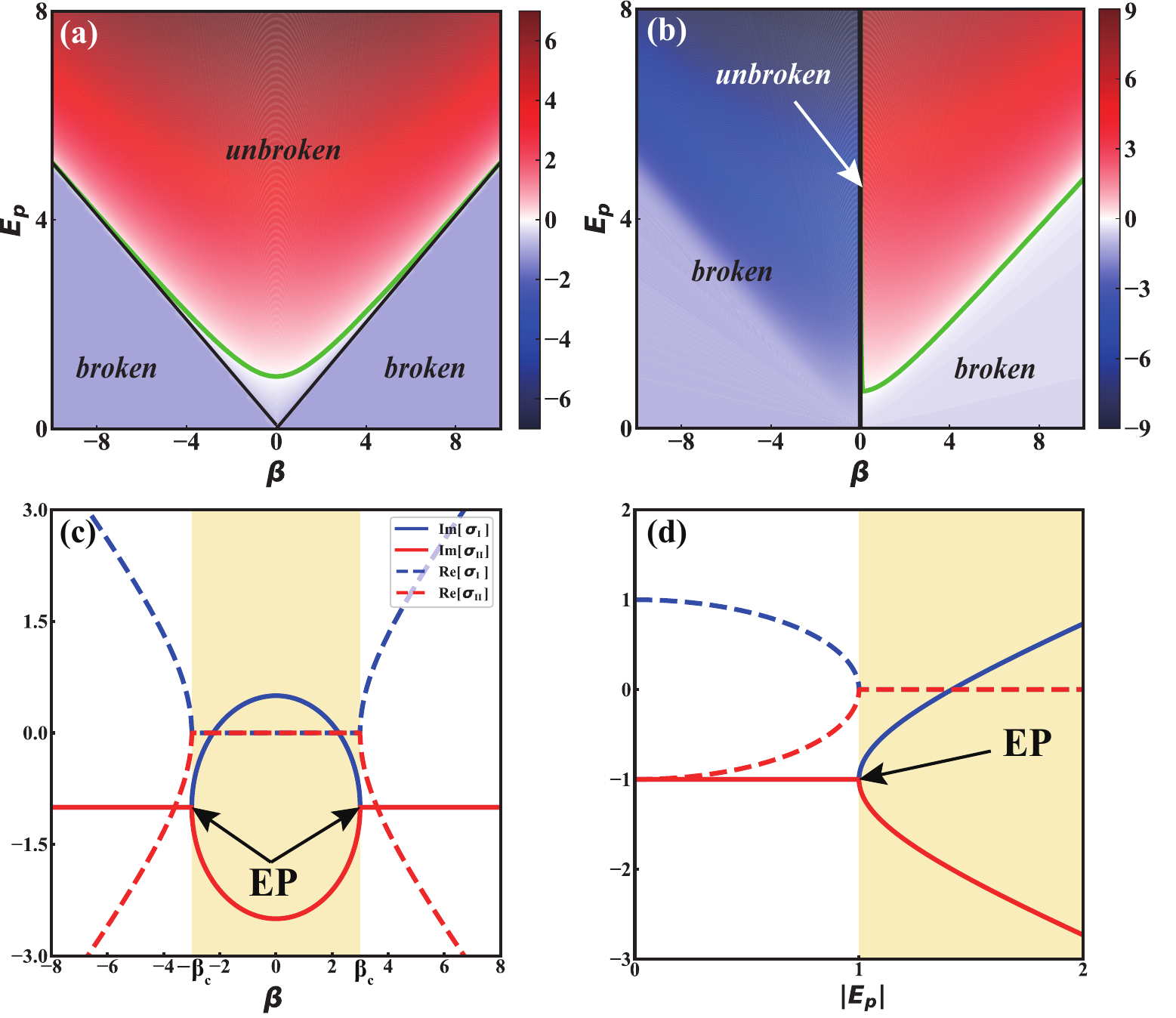}
	\caption{The amplification/decay of (a) the gain eigenmode for systems with symmetric absorption ($\gamma_s=\gamma_i=1$) and (b) one of the eigenmode for systems with asymmetric absorption ($\gamma_s=0.5,\gamma_i=1$). The black and green curves represent the boundary of anti-PT symmetry and amplification. (c,d) The presence of EPs considering symmetric absorption ($\gamma_s=\gamma_i=1$) for (c) $|E_p|=1$ and (d) $\beta=1$. For all the plots, $\kappa_s=\kappa_i=1$.}
	\label{fig:3}
\end{figure}

\textit{Anomalous behaviors for eigenmodes in unbroken and broken anti-PT symmetric regimes}--Anomalous behaviors for eigenmodes can be observed in amplification regimes irrespective of anti-PT symmetry. For simplification, in the case of symmetric absorption for signal and idler by fixing $\gamma_s=\gamma_i=1$, the calculated imaginary parts of eigenvalues for the amplified eigenmode $\rm{Im}(\sigma_+)$ as shown in Fig.~\ref{fig:3}(a) reveals that, $\rm{Im}(\sigma_+)$ is symmetric regarding $\beta$. Furthermore, for anti-PT symmetric systems, one eigenmode is always amplified while the other is always decayed. However, for anti-PT-symmetry-broken systems with symmetric absorption for signal and idler photons, both the eigenmodes always decay. In the case of asymmetric absorption for signal and idler by setting $\gamma_s=0.5$ and $\gamma_i=1$ belonging to the fourth quadrant shown in Fig.~\ref{fig:1}(a), the calculated $\rm{Im}(\sigma)$ for one of the eigenmodes is shown in Fig.~\ref{fig:3}(b), which reveals that the amplification/decay of two eigenmodes is odd regarding the sign of the phase mismatch $\beta$, i.e. amplification for positive $\beta$ while decay for negative $\beta$. Such gain/loss-sgn($\beta$) locking phenomenon only occurs in asymmetric-absorption systems, which has been rarely observed in conventional OPA systems.

\textit{Exceptional points}--The two eigenvalues of Eq.~(\ref{equation:1}) are degenerate for systems with symmetric absorptions of $\gamma_i=\gamma_s$ and the phase mismatch satisfying $\beta=\pm\beta_c$. In this condition, the eigenstates of the system also collapse, which indicates exceptional point (EP) appears as shown in Fig.~\ref{fig:3}(c). EPs appear in pairs with positive and negative phase mismatch, which are also the transition point separating unbroken and broken anti-PT symmetry regions. When the absorption $\gamma$ changes, the EP changes along the blue solid line as shown in Fig.~\ref{fig:1}(e), therefore, the systems in the $\beta-\gamma$ parameter space possess two exceptional lines. For systems with fixed $\beta$ and $\gamma$, the pump intensity changes following the change of $\beta_c$, thus only one EP can be observed in $|E_p|$ parameter space as shown in Fig.~\ref{fig:3}(d). According to the simulations of phase-mismatch $\beta-$dependent signal-gain evolution, a slight signal-gain oscillation exists in the large $\beta$ region, while there is a jump in signal gain can be observed when $\beta$ passes the EP threshold value, as shown in Fig.~\ref{fig:s2} and Fig.~\ref{fig:s3}. The anomalous behaviors around the EPs in non-Hermitian OPA systems are distinctly different from those in conventional non-Hermitian optics systems, which have been intensely investigated for their super sensitivity to any external perturbation\cite{ElGanainy2019}. See detailed discussions on the anomalous parametric gain behaviors around EPs/ELs in non-Hermitian OPA systems in SUPPLEMENTARY NOTE 6.

\begin{figure}[htbp]
	\centering
	\includegraphics[width=1\linewidth]{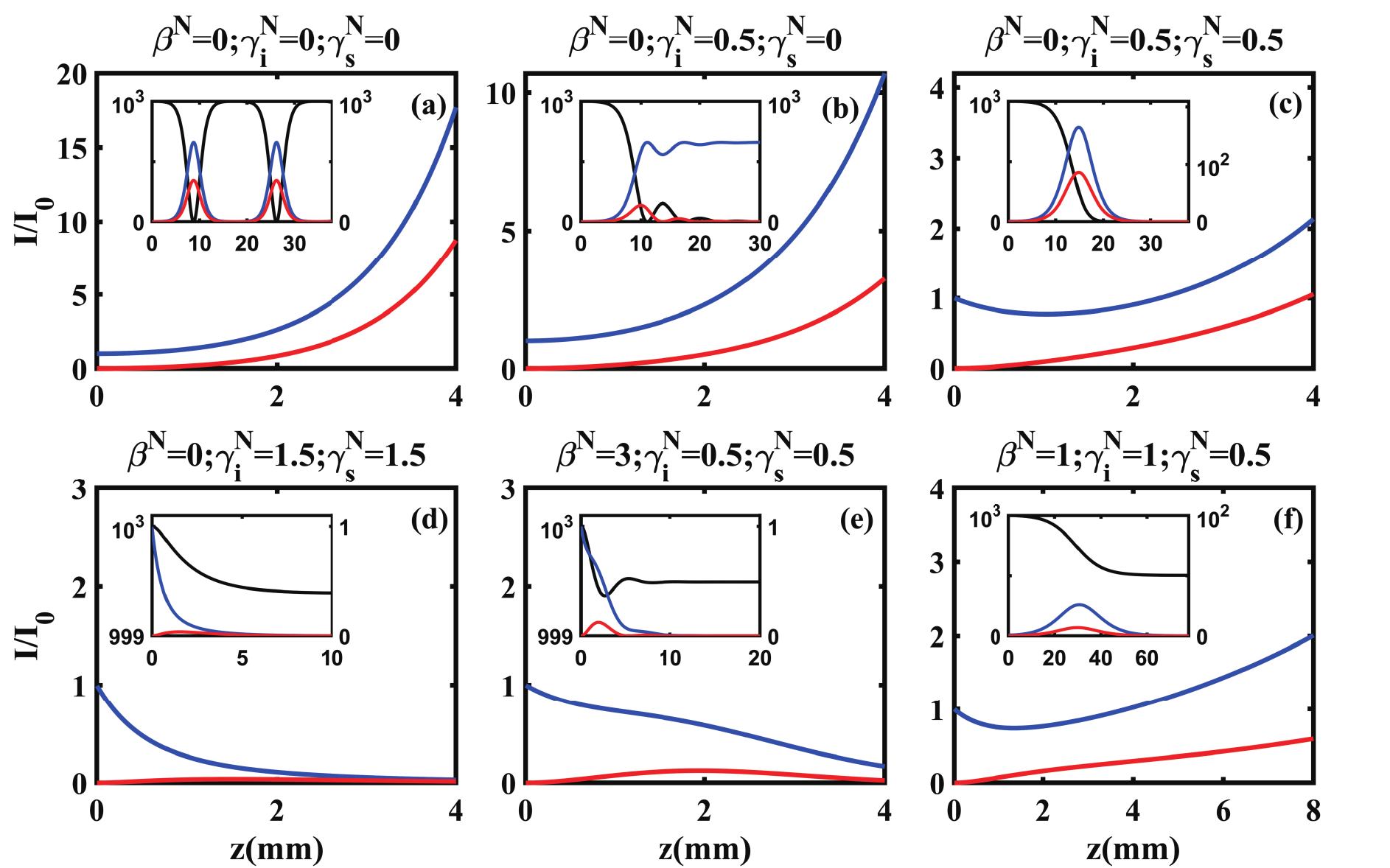}
	\caption{Evolution of the intensity of the signal (blue curve) and the idler (red curve) beams as a function of the crystal length. Case under phase matching condition $\beta^N=0$, (a) OPA progress, $\gamma_{i,s}^N=0$, (b) QPA progress, $\gamma_i^N=0.2$, $\gamma_s^N=0$; and symmetric absorptions for (c)$\gamma_{i,s}^N=0.5$, (d) $\gamma_{i,s}^N=1.5$. Case under phase mismatching condition (e)$\beta^N=3$ and $\gamma_{i,s}^N=0.5$, (f)$\beta^N=1$, $\gamma_i^N=1$ and $\gamma_s^N=0.5$. Insets in (a)-(f) are the evolution of the intensity to the saturation region, signal and idler correspond to the right axis scale, and pump to the left axis scale.}
	\label{fig:4}
\end{figure}

\textit{Simulations for practical systems}--We firstly scale $eigs(\mathcal{H})$ by $\gamma^N_{i,s}=\gamma_{i,s}/\Gamma$, $\beta^N=\beta/\Gamma$, and $\sigma^N_{\pm}=\sigma_{\pm}/\Gamma$, with $\Gamma=\sqrt{\kappa_s\kappa_i|E_{p0}|^2}$ referring to the nonlinear drive and $E_{p0}$ the initial pump amplitude. Fig.~\ref{fig:4} depicts the simulated evolution of signal and idler intensities $I_s$ and $I_i$ as a function of the crystal length $z$ within the small-signal and saturation (insets) regions. For conventional phase-matching OPA systems without absorption ($\gamma_{s,i}^N=0$) and phase-matching QPA systems with small idler absorption ($\gamma^N_s=0, \gamma_{i}^N=0.2$), both the signal and idler exhibit purely similar exponential amplification within small signal region, as shown in Fig.~\ref{fig:4}(a,b). When considering the saturation region with long crystals, the signal and idler intensities experience conversion-back-conversion cycles and reach peak values when the pump is depleted for conventional OPA systems, while the energy-flow cycle is broken for QPA systems. The reason is that part of the idler photons are absorbed, which breaks the duality of the signal-idler photon pairs in terms of their quantities. The imbalance in the number of signal photons and idler photons prohibits the energy flow from the signal to the pump when the idler photons are depleted.

For the phase-matching ($\beta^N=0$) case with symmetric absorptions for signal and idler photons of $\gamma_{s,i}^N=0.5$ in the first quadrant denoted by the star in Fig.~\ref{fig:1}(e), the system is in the anti-PT-symmetry regime with amplification, and the signal exhibits weak decrease before the exponential increase with $z$ as shown in Fig.~\ref{fig:4}(c). The decrease is due to the nontrivial absorption of signal photons which exceeds the generation of signal photons via the conversion process from pump photons in the early stage. A similar decrease can be observed in the other regimes shown in Fig.~\ref{fig:4}(f). When the symmetric absorptions increase to $\gamma_{s,i}^N=1.5$ in the second quadrant denoted by the hexagon in Fig.~\ref{fig:1}(e), the system is in the anti-PT-symmetry regime with decay. The signal exhibits purely exponential decrease over the whole crystal length $z$ as shown in Fig.~\ref{fig:4}(d). When a large phase mismatch of $\beta^N=3.0$ ($\beta^N>\beta_c^N=2.0$) is considered in systems with small symmetric absorptions of  $\gamma_{s,i}^N=0.5$ denoted by the triangle in Fig.~\ref{fig:1}(e), it is in the anti-PT-symmetry-broken regime with decay, which belongs to the third quadrant. The signal exhibits a rapid decrease with $z$ in a mildly fluctuating manner. For a system with asymmetric absorptions of $\gamma_s^N=0.5, \gamma_i^N=1.0$ and a small phase mismatch of $\beta^N=1.0$ denoted by the ball in Fig.~\ref{fig:1}(f), it is in the anti-PT-symmetry-broken regime with amplification, belonging to the fourth quadrant. The evolution of the signal beam on $z$ is similar to that in the first quadrant, except for its faster increase rate on $z$ due to the larger absorption of idler photons.

As the crystal becomes long enough, for the system in the anti-PT-symmetry regime with amplification, the signal and idler intensities reach peak values before the pump is depleted, as shown in the inset of Fig.~\ref{fig:4}(c). As the pump intensity decreases, the rate at which the pump energy flows into the signal and idler is less than the rate at which the signal and idler are dissipated, resulting in the energy of the three beams decreasing until they are depleted. A similar evolution occurs in the system in the anti-PT-symmetry-broken regime with amplification, where the difference is that the pump is not completely depleted, as shown in the inset of Fig.~\ref{fig:4}(f).

\textit{Discussion}--As shown in Fig.~\ref{fig:1}(e), the signal amplification-decay boundary ($|\beta|<\sqrt{\beta_c^2-4\gamma^2}$) restricts the signal amplification bandwidth, known as the phase-matching bandwidth in the OPA process previously. The non-Hermitian anti-PT symmetric system created by introducing dissipation, has the potential to fundamentally broaden the signal amplification border, as shown from the analysis of parameter space in this study. In other words, the aforementioned signal amplification boundary is broken, such that the signal gain is no longer limited by the phase mismatch, manifesting the potential of meta-broadband amplification. High-efficiency and broadband amplification in the near infrared wavelengths has been accomplished experimentally\cite{Ma2015,Ma2017}. The crucial use of non-Hermitian systems in enhancing mid-infrared broadband amplification efficiency is still being investigated in this field. Moreover, the mode switching near the EP enables ultrafast gain switching through pump-controlled breaking of PT symmetry, which has been proved to be a new method for the direct generation of an ultrashort pulse train in PT symmetric coupled OPA waveguides\cite{Wang2023}. Similar effects are expected to exist in single crystals and various strategies for the non-linearity coupling to different anti-PT symmetric optical structures to increase the capabilities of optical manipulation are to be proposed.

\textit{Conclusion}--In summary, we have investigated the dynamical behavior of non-Hermitian parametric amplifiers based on anti-PT symmetry in a single crystal. From the Hamiltonian model built from the coupled-model equations for three-wave interactions, the anti-PT symmetry operation is introduced to restrict special eigenvalues of OPA systems. By considering the parameter space formed by the phase match $\beta$, absorptions for signal and idler channels, $\gamma_s$ and $\gamma_i$, the full-landscape behaviors of the non-Hermitian OPA systems were categorized into four quadrants regarding the unbroken/broken anti-PT symmetry and amplification/decay. We also simulated the evolutions of signal and idler beams in systems belonging to the four quadrants respectively and demonstrated their rich dynamical behaviors. This work not only accounts for the conventional OPA and the recently proposed QPA systems, but also may inspire novel designs beneficial for further enhancement of the efficiency and the amplification bandwidth.

\vspace{15mm}
\bibliography{OPA}

\end{document}